# FUTURE TRENDS AND CHALLENGES FOR MOBILE AND CONVERGENT NETWORKS.

*José André Moura*[*]
ISCTE-IUL, Instituto Universitário de Lisboa
IT, Instituto de Telecomunicações, Portugal
*Christopher Edwards*[*]
Lancaster University, Infolab21, UK

## ABSTRACT

Some traffic characteristics like real-time, location-based, and community-inspired, as well as the exponential increase on the data traffic in mobile networks, are challenging the academia and standardization communities to manage these networks in completely novel and intelligent ways; otherwise, current network infrastructures can not offer a connection service with an acceptable quality for both emergent traffic demand and application requisites. In this way, a very relevant research problem that needs to be addressed is how a heterogeneous wireless access infrastructure should be controlled to offer a network access with a proper level of quality for diverse flows ending at multi-mode devices in mobile scenarios.

The current chapter reviews recent research and standardization work developed under the most used wireless access technologies and mobile access proposals. It comprehensively outlines the impact on the deployment of those technologies in future networking environments, not only on the network performance but also in how the most important requirements of several relevant players, such as, content providers, network operators, and users/terminals can be addressed. Finally, the chapter concludes referring the most notable aspects in how the environment of future networks are expected to evolve like technology convergence, service convergence, terminal convergence, market convergence, environmental awareness, energy-efficiency, self-organized and intelligent infrastructure, as well as the most important functional requisites to be addressed through that infrastructure such as flow mobility, data offloading, load balancing and vertical multihoming.

**Keywords**: Convergence, Flow Mobility, Load Balancing, Wireless Access, Future Networks, Multihoming

---

[*] Corresponding Author address
    Email: jose.moura@iscte.pt
[*] Corresponding Author address
    Email: c.edwards@lancaster.ac.uk

# INTRODUCTION

Mobile users using handheld devices require services in a similar way as they had previously using the wired network, such as, video streaming or IPTV. In addition, a global rollout of a single radio access technology is not foreseen due to several reasons: i) the existence of several wireless technologies and flow requirements, ii) the inexistence of available spectrum, iii) network operators protecting their previous large investments and iv) frequent network congestions due to the current high popularity of more powerful handheld multimode devices.

It is pertinent to remember the vision of International Telecommunication Union - Telecommunication Standardization Sector (ITU-T) for future networking environments. This vision is namely about the design of general signaling system that can allow the convergence (interworking) of diverse network access technologies (e.g., cellular networks, broadband networks, wireless LANs) in to a single IP network infrastructure, i.e. Next Generation Network (NGN) [29]. The specific requisites that each access technology should satisfy to enable the formation of a global NGN infrastructure are normally designated by 4G (and beyond) requisites or IMT-Advanced requirements [110]. Some examples of 4G requisites are interoperability with existing wireless standards, seamless connectivity across multiple networks, and ability to offer high end-to-end quality of service for multimedia traffic.

Future networking environments will incorporate most likely, simultaneous usage of multiple access networks because the usage of smartphone terminals with multiple wireless interfaces is increasing exponentially. All these upcoming capabilities can catalyze the discover of intelligent ways to manage the complete set of resources available from the diverse interfaces (technologies) in such way the entire resources can be used to support users' connection quality in scenarios with a significant amount of data traffic. A first example of intelligent networking management without compromising the users' connection quality is the cooperative usage among technologies of networking resources [58]. Other illustrative example is the offloading of data traffic from an overloaded technology to alternative lighter technologies. In fact, to reinforce the relevance of this last example, a recent study has concluded that 33 percent of total mobile data traffic was offloaded onto the fixed network through Wifi or femtocells [78]. In addition, some popular applications in the next couple of years will be interactive video and also these will offer customized data based on, as an example, the user's location, the terminal being used and/or user's profile.

Assuming that network access technologies would converge into a unique abstracted cloud of wireless coverage, the users expect to have a good quality access to a large diversity of services, independently of their location and mobility behavior, through the usage of multimode and handheld devices. In this way, the network management should be revisited to use efficiently the available resources of heterogeneous wireless access networks in an adaptive, cooperative and integrated way. Some relevant aspects that urged to be revisited are related to how the wireless access and user mobility are currently supported and how they can be enhanced to support, namely, the data growth on the traffic data without disrupting the network operation and the fulfilling of emerging applications' requisites.

The current chapter is organized as follows:

• It initially summarizes and refreshes the current literature [65,67-70,73,77,79-82,86-89,90,91,105-107] to identify the more suitable solutions to deploy and manage the mobile access in future networks where the available applications (services) are clearly splitted off underneath the transport infrastructure [111] [112];

•    All the selected mobile access technologies are comprehensively discussed along the current document following a methodology that can be decomposed in three parts. The first part briefly presents the technology in a tutorial manner. The second part points out the strengths and drawbacks of that technology over a set of relevant requirements for future networks, according to the view of different players: users/terminals, network and service providers. The third part discusses some potential issues associated to the deployment of that technology in future networks;

•    The final part of the current chapter highlights our major conclusions and future developments in the mobile access to future networks.

# LITERATURE REVIEW

This section revises some very relevant topics in future networks, as follows:
•    Convergence on wireless access technologies;
•    Support of mobile access.

## Convergence on the Future Wireless Access

As the network resources of point-to-point links in heterogeneous networks, with the increasing on data traffic [78], approach their own maximum capacity limits, researchers have developed a big effort to find effective proposals to manage the access to emergent Next Generation Network (NGN) environments in a way that the new challenging requirements could be efficiently satisfied with the available network resources [74]. At the time of this writing, the mobile operators are dealing with a very important issue due to the lack of capacity in their network infrastructure, originally dimensioned to support only voice and messages. In this way, several congestion situations have been reported, disrupting the functionality of mobile cellular networks because these cannot support the huge increase on data traffic. The mobile operators have pinned out some immediately solutions to counteract the congestion issue, such as, i) upgrades on the network infrastructure, ii) offloading traffic. The authors of the current work argue that the congestion problem can be also avoided by managing intelligently the complete set of network connectivity resources from all the available wireless access technologies in a specific geographic location, supposing a convergence scenario among these technologies [97]. Some good hints into how to perform this can be found in [64]. It covers comprehensively different aspects of analysis, design, deployment, as well as some optimization techniques to be applied to protocols and architectures for heterogeneous wireless access networks. In particular, the discussed topics are namely the following: convergence of distinct access networks, cognitive techniques to manage radio resources, admission control and network selection, energy efficiency, pricing and content discovery. Other contribution [57], comprehensively reviews Vertical Handover Decision (VHD) algorithms grouping these into four distinct categories. This taxonomy is based on the most important decision parameter used by each VHD algorithm to select the more convenient network/NAP, main decision used by each of these discussed algorithms before issuing a handover, such as [100]: i) RSS-based, ii) bandwidth-based, iii) cost function-based and, iv) combination of several parameters algorithms. An alternative VHD taxonomy can be found in [103], and a related revision work more focused on VHD over LTE-Advanced is available in [104]. The authors of [97] discuss comprehensively the convergence of IMT-Advanced access networks (LTE, Wimax) including ITU-R requirements, such as, new frame structure, spectrum operation and, supporting the increasing demand for mobile data.

Spectral efficiency of cellular networks is thoroughly investigated in [71] to enhance the network performance. They discuss LTE scenarios using a deployment strategy that coordinates the interference and balances the load among the network nodes. Aligned in the same direction line, the authors of [72] believe into a rapid acceleration towards femtocells. They also discuss some related pertinent issues: deployment coordination, cost impact and eventual chaos introduced in the normal network operation by the introduction of femtocells.

While a variety of Multiple-Input and Multiple-Output (MIMO) receiving/transmission techniques are available at the BS/eNodeB side to enhance network capacity, the terminal options to use MIMO are currently very limited. In fact, there is a technological limitation imposed by the maximum number of antennas supported by a single terminal. The available terminal options are cooperative diversity, dedicated relay stations, and femtocells, which are further discussed and compared in [97]. In addition, a very recent technique designated by Coordinated Multi-Point (CoMP) transmission exploits rather than mitigates inter-cell interference, enhancing the network throughput at the cell-edge [97].

A very important functional aspect is to ensure an efficient and seamless roaming across a NGN environment, through a sequence of some well identified phases, e.g. the following three phases. The first phase is related with link-layer network discovery in a technology-dependent way for horizontal handovers [52] or technology-agnostic way for vertical handovers [9]. The second phase is selection of the more suitable access network/Network Attachment Point (NAP) [49]. The third phase is about predicting handoffs [83] to enhance the handover/handoff management in currently deployed mobile networks [84].

The current standardization bodies are very focused in seamless heterogeneous handoffs [4] [65]. In this way, the authors of [85] propose a more holistic vision to support mobility in the highly complex NGN environment through cognitive handoffs, which are controlled by a variety of policies, and can attain multiple purposes simultaneously.

The usage of complementary wireless technologies to enhance the mobile cellular networks currently deployed would create significant changes in the business models for mobile telecommunications industry [50]. Nevertheless, some aspects can impair the innovation in this area [51], like the availability of RF spectrum.

Finally, one should be aware that the development of future networks should involve not only the access technology convergence discussed before but also other convergent aspects [112], namely the following ones: the same information is offered through different platforms, laptops and smartphones are quickly converging to a single type of terminal, and both telecom operators and broadcast providers are selling similar «n-play» service packages.

## Support of Mobile Network Access

A considerable number of networking publications discussing the support of mobility has been found in the literature. From all these, the more recent and high-quality contributions have been selected. In addition, the same contributions have been classified according to their main topics, as shown in Table 1.

An interesting tutorial can be found in [105] about mobility management in data networks, namely on session migration. In addition, some surveys related with IETF mobility proposals have been also found, covering the following aspects: macro mobility [65], route optimization in network mobility (NEMO) [69], IPv6 multihoming solutions [80], and ID/Locator split architectures [82].

The theme of mobility support in heterogeneous access networks is covered in [73][77][80][81][90][91]. From these, references [81][90][91] discuss more particularly multihoming-based solutions but in distinct ways. For example, [90] discusses multi-homing in a broader context, with criteria like robustness, ubiquitous access, load and flow management. Alternatively; [91] restricts its study to an unique criterion like cost. In addition, reference [73] studies proposals not only focused in multihoming but also covering mobile scenarios.

Table 1: Literature survey

| Main topic | References |
|---|---|
| Session migration (tutorial style) | [105] |
| Mobility management | [65][73] |
| Network mobility (NEMO) | [67][69][70] |
| Heterogeneous access networks | [73][77][80][81][90][91] |
| Multihoming | [73][81][90][91] |
| ID/Locator split | [82] |
| Offloading | [87][95][106] |
| Train networks | [88] |
| Vehicular networks | [67][68][79] |
| Aeronautical networks | [70] |
| Satellite networks | [89][107] |

The authors of [86] initially discuss the on-going evolution in the 3rd Generation Partnership Project (3GPP) architectures to flat and ultra-flat designs trying to scale up their performance and satisfy the continuously growing traffic data demands. Then, they proceed with the discussion in how those flat mobile Internet architectures can efficiently support distributed mobility management schemes. They finalize their contribution summarizing the challenges to interconnect the future mobile flat architectures to the Internet. Some offloading strategies are also discussed in [106] [87] [95] to counteract the overloading problem in mobile networks. The deployment of femtocells can further reduce the congestion by offloading traffic to ADSL cable/optical fiber. In this way, the handover of flows could be initiated by either host mobility or available capacity of networks to satisfy traffic requirements/user preferences.

A recent contribution [88] discusses how to offer a reliable Internet access to passengers on trains. This solution can use 802.11 as the access technology within the train and cellular/Wimax/satellite as possible backhaul technologies.

The authors of [67] provide a qualitative evaluation among diverse IP mobility solutions that enable NEtwork MObility (NEMO) [19] routing in vehicular networks. The discussed requirements are energy-efficient transmission at terminals, reduced handover events, lower complexity, reduced bandwidth consumption, minimum signaling, security, binding privacy protection, multihoming, and switching Home Agent (HA). Further surveys in vehicular networks are available in [68] [79]. They outline a significant list of open research issues: access selection, mobility model, ad hoc routing, handover latency reduction vs. QoS provisioning, vehicle mobility modeling, non-uniform access network coverage scenarios, and security issues.

The authors of [70] also propose NEMO as the more convenient mobility solution to aeronautical communications. They conclude identifying pertinent future work, as follows: the synchronization among all home agents about the location information of mobile nodes, route optimization and end-to-end latency are not properly addressed, as well as dealing properly with the packet loss issue.

Mobility support in satellite networks is covered in [89][107]. Satellite networks can be a viable option to cover remote areas with no available network infrastructure.

From previous research work, such as [32-33], we can envisage that the best way to support mobility is still an open issue, which requires further research and standardization effort.

## FUTURE NETWORK REQUIREMENTS

This section describes the main functional requirements that, according to the opinion of the authors of the current work, any mobile access technology used in future networks (i.e. NGN environment implemented over a 4G heterogeneous network infrastructure) should satisfy. These nine requirements are listed in Table 2 and are following explained.

Table 2 – Future Network Requirements

| Id | Description |
|----|-------------|
| R1 | Message forwarding |
| R2 | Route update |
| R3 | Handover efficiency |
| R4 | Mobile node location |
| R5 | Security |
| R6 | Robustness |
| R7 | Concurrent movement |
| R8 | Deployment |
| R9 | Scalability |

The first requirement is designated by message forwarding. It is about the successful delivery of messages to final destinations in spite of the eventual disruption originated by handovers (or handoffs). These handovers, traditionally justified by node mobility, in the future network environment, can also be due to the dynamic selection of an alternative Network Attachment Point (NAP) with a higher connectivity quality than the one being used. This means static nodes can also perform a handover if, for example, a management policy to increase the connection quality decides to move a terminal to an alternative NAP offering a better network access service.

The second requirement is the route update that characterizes how fast a new routing path is propagated across the network, including the mobility agents or correspondent nodes, after a node has moved to other NAP. Ideally, the packets should be delivered with success to their final destinations in spite of these being mobile terminals.

The third requirement is concerned with how efficiently the technology manages handovers, minimizing packet loss, network overhead and delay. In this way, the handover process should not disrupt the quality associated to traffic flows used by the mobile terminal. This requirement is more pertinent for long-time flows than for short-time ones. In fact, the effect induced by, for example, a handover process with high latency could be almost unnoticeable by a short-time flow due to the fact that flow during its short existence did not need any handover.

The fourth requirement is how often a terminal becomes active from its dormant state. There is a tradeoff between the battery autonomy and how fast the terminal is moving between cells. If the terminal is changing between cells very often then that terminal is required to perform its cell registration also very often. In this case, the terminal battery energy can become exhausted very quickly. This situation becomes worst as the terminal is multimode.

The fifth requirement is related with security. The mobility management solution should not introduce any new security vulnerabilities. As an example, the client privacy should be always guaranteed. In this way, the location of user terminals should not be exposed to malicious nodes nor exposed to service providers (i.e. Correspondent Nodes - CNs).

The sixth requirement is related with resiliency and robustness. In fact, the management mobility protocol should be robust against any network failure. If the terminal is multimode and there are multiple access technologies at a specific location, then these two facts can enhance the robustness against a failure on a wireless access technology, as the mobility protocol after detecting that failure can move flows from the faulty technology to the other one.

The seventh requirement is related with the fact that the management mobility protocol should operate perfectly well in spite of the concurrent movement of both mobile node and CNs, which is a very plausible scenario because some mobile users (i.e. mobile CNs) can make multimedia contents directly available to others.

The eighth requirement is associated with the cost for deploying a new management mobility protocol, which should be evaluated from the point of view of network provider. In addition, novel deployments should be transparent to the end-user, avoiding any upgrade on terminal/software.

The ninth requirement evaluates how well a management mobility protocol scales in terms of the number of mobile and correspondent nodes. The management proposal should also support a high ratio of handover requests without disrupting the flows.

Our chapter analyzes some selected mobile technologies, which are listed in Table3. To proceed with the previous analysis, we also be using the future network requirements previously discussed in the beginning of the current section and summarized in Table4. From this table, one can easily identify how the different system players are sensitive to those requirements. It is assumed the terminal and the user are different players in order to accommodate their distinct requisites (autonomy vs. privacy). From Table 4, one can also conclude that requirements R1, R2, R3 and R7 (described in Table 2) are essential to efficiently support mobility in future networking environments. In this way, these requirements normally affect all the involved players: service providers (i.e. CNs), network providers (e.g. signaling overhead/delay associated to route update), terminals (e.g. handover delay) and users (e.g. QoE - Quality of Experience).

The requirements R6, R8 and R9 have a strong impact on the deployment of the network infrastructure. Consequently, these requirements essentially affect network providers (e.g. robustness, deployment factors, scalability aspects) and terminals (e.g. deployment factors).

The requirement R4 has a strong effect on the terminal battery autonomy. Finally, requirement R5 has a relevant influence on the user privacy.

Table 3 – Mobility protocols under Study

| Layer | Proposals |
|-------|-----------|
| 3 | MIPv4, MIPv6, HMIPv6, FMIPv6 |
| 2 | IEEE 802.11, 3GPP LTE, IEEE 802.16 |

Table 4 – Which players can be affected by fulfilling future network requirements

| Requirements | Main Goal | Affected Players |
|--------------|-----------|------------------|
| R1, R2, R3, R7 | Mobility Support | All |
| R6, R8, R9 | Deployment | Network Provider, Terminal |
| R4 | Autonomy | Terminal |
| R5 | Privacy | User |

# WIRELESS ACCESS TECHNOLOGIES

The current section discusses the more relevant existing and emerging wireless access technologies: IEEE 802.11, 3rd Generation Partnership Project (3GPP) LTE and IEEE 802.16. The latter two technologies have been recently recognized as IMT-Advanced technologies [97], and the former one is massively deployed in public hotspots and residential areas. In this way, 802.11 can enhance the access connectivity provided by either LTE-Advanced or 802.16m, for example, by mitigating congestion through the offloading of data traffic from the network core to the 802.11 infrastructure. Each of the following access technology is systematically presented, discussed, and analyzed in the next three perspectives: how the technology evolved and its expected evolution (i.e. background), how the technology operation can affect (positively or negatively) the players expectations discussed in the previous section (i.e. critical analysis), and some potential issues associated to the practical deployment of that technology.

## IEEE 802.11(WIFI)

Background

This sub-section aims to discuss in a tutorial style the most important layer 2 functional mechanisms to enable IEEE 802.11 compatible equipment to be incorporated in a heterogeneous network access infrastructure, which offers a broadband mobile connection.

The IEEE 802.11 equipment has had a widespread deployment due to its low cost. In addition, several mobile operators offer contracts to their customers, enabling them to join a vast number of wireless hotspots available worldwide [47][98], at diverse locations such as airports, railway stations, malls, university campuses and convention centers. In this way, the basic cellular coverage can be enhanced with the additional one provided by wifi when available and, assuming the terminals support multiple access technologies. This coverage enhancement is made possible because there are also business (roaming) agreements among mobile operators and Wifi network providers [99]. There is also the case of a worldwide Wifi coverage such as the one offered inside the FON community [98]. Unlike mobile cellular access, public WLAN networks can offer higher bit access rate to the Internet such as 54 Mb/s [3], or eventually higher rates. These WLANs utilize either 5 or 2 GHz spectrum bands that are publicly available.

In the following text, we briefly explain how 802.11 operates at layer 2. The 802.11 MAC sublayer is responsible for several functions, namely to coordinate the multiple access to the same radio channel. This channel access can be done in two possible modes as visualized in Figure 1: a decentralized one designated by Contention Period (CP), and a centralized one designated by Contention-Free Period (CFP). During the CFP, the AP pools sequentially the data from each station. The CFP mode initiates right after a Beacon message is disseminated by the AP within the Wifi cell. Alternatively, during the CP mode, the stations contend for the channel access before each packet transmission through the Distributed Coordination Function (DCF).

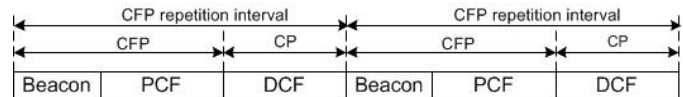

Figure 1 - IEEE 802.11 Modes

Each station transmitting via CP mode uses the channel access algorithm designated by Carrier Sense Multiple Access with Collision Avoidance (CSMA/CA). In CSMA/CA, a station before sending a frame checks the channel status. As the channel is "idle", the station transmits right away. Alternatively, the station defers transmission for a "random" (i.e. backoff) interval. This last behavior tries to reduce the probability for collisions. A collision occurs when several stations transmit simultaneously on the same radio channel.

The 802.11 DCF only provides a best-effort channel access to all the available traffic types. In opposition, the 802.11e EDCF introduces priority traffic differentiation, to fulfill the distinct QoS requisites of several application types [66] [61]. A single 802.11e node has four frame queues, one queue for each Access Category (AC), as shown in Figure 2. In addition, this node has also four EDCF channel access functions, one for each AC queue. So, the EDCF mode tries to initially schedule the channel access among four traffic types, e.g., background, best effort, video and voice, at distinct time intervals but with a well defined priority. In this way, it is giving to voice traffic the highest priority in terms of channel access; and to backoff traffic the lowest priority. To implement this differentiation, each AC has a distinct setup, i.e. AIFS[AC]. The AC with the highest priority normally is configured with the lowest value on its AIFS[AC] parameter. Then, in a second step, the EDCF tries to avoid collisions among contending flows of the same AC using a backoff mechanism configured by CWmin[AC] and CWmax[AC] parameters that are both used to evaluate the CW[AC] window. This window is used to randomly select the backoff timer among the several contending flows (from distinct stations) belonging to the same AC. In the case of successive collisions in the channel the backoff timer is (exponentially) enlarged to enable the colliding stations to transmit their frames, a single one of these each time. Otherwise, the backoff has a minimum value to offer a fast channel access.

Table 5 gives an overview of the IEEE 802.11 research effort at the time of this writing, which seems a very dynamic working area. The current IEEE 802.11 standard was approved in 2012. Late 2015 is expected to be approved a new 802.11 standard merging and incorporating, at that time, some approved amendments (e.g. 802.11aa/ac/ad). In addition, other task groups are currently working towards a final amendment approval (e.g. IEEE 802.11ax) for further increasing the efficiency of WLANs.

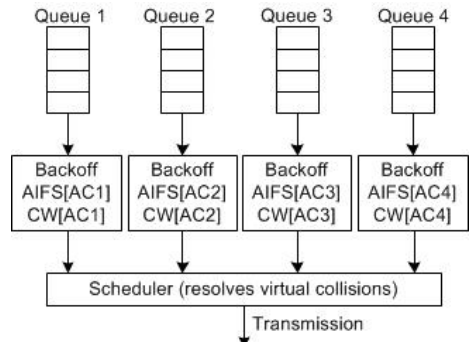

Figure 2 - IEEE 802.11e EDCF channel access

Table 5 – IEEE 802.11 Standardization Updated Perspective

| Standard/Amendment | Responsibility |
|---|---|
| 802.11-2007 | IEEE standard that includes the previous following amendments: a, b, d, e, g, h, i and j |
| 802.11-2012 | IEEE standard that merged ten amendments (802.11k, r, y, n, w, p, z, v, u, s) with the 2007 base standard |
| 802.11aa-2012 | It is an approved amendment that specifies enhancements to the 802.11 MAC for reliable audio/video streaming |
| 802.11ac-2013 | It is an approved amendment to IEEE 802.11, published in December 2013, that builds on 802.11n (MIMO optimizations) |
| 802.11ad-2012 | It is an approved amendment to IEEE 802.1, published in January 2013, that provides data rates up to 7 Gbps |
| 802.11af-2014 | It is an approved amendment to IEEE 802.1, published in February 2014, that through cognitive radio technology allows WLAN operation in TV white space spectrum in the VHF and UHF bands between 54 and 790 MHz |
| IEEE 802.11ah | Upcoming amendment (expected in 2016) to define WLANs operating at sub 1 GHz license-exempt bands. This could allow traffic offloading from mobile networks. |
| 802.11ai | Upcoming amendment to standardize a fast initial link setup function that would enable a wireless LAN client to achieve a secure link setup within 100ms. |
| 802.11aj | Upcoming amendment that rebands of 802.11ad for use in the 45 GHz unlicensed spectrum available in some regions of the world (specifically China) |
| 802.11aq | Upcoming amendment to the 802.11 standard that will enable pre-association discovery of services. This extends some of the mechanisms in 802.11u that enabled device discovery to further discover the services running on a device, or provided by a network |
| 802.11ax | Upcoming amendment such as a successor of 802.11ac for increasing the efficiency of WLANs. The initial goal of this project is to increase fourfold the throughput of 802.11ac |

<u>Critical View</u>

This sub-section discusses some research efforts to enhance 802.11 networks, which can have a significant impact on future network requirements. All these found research contributions are summarized in Table 6.

The initial main goal of WLANs was to diminish the cost to deploy a cabled infrastructure in indoor environments typically using applications requiring low data rates. In addition, the mobility support initially was not a critical requisite. However, the mobility has turned on a pertinent requisite to be fulfilled by WLANs mainly if these are within a heterogeneous wireless broadband infrastructure. Consequently, enhancing the handoff support in 802.11 networks has become a very important functional aspect to provide mobile services, as well as the support of applications requiring higher data throughput and low latency/jitter. The 802.11K working group (WG), using radio resource measurement, studies how to produce and disseminate meaningful reports to MNs listing the entire set of candidate APs covering a specific area. Then, the MN applies to the received AP lists some ranking criteria to choose the more convenient AP as the next Network Attachment Point (NAP). The main mobile requirements addressed by this proposal are R1 (message forwarding) and R3 (handover efficiency). Related work is available in [52]. They propose an algorithm to diminish the computing complexity and effort to model the area covered by a grid of APs.

Table 6 – Future network mobile requirements addressed by IEEE 802.11 technology

| Addressed Requirements (Positive Contribution) | Player | References |
|---|---|---|
| R1 (Message forwarding), R3 (Handover efficiency) | All | IEEE 802.11K, [52], 802.11F |
| R2 (Route update) | All | IETF DNA [5], [38] |
| R3 (Handover efficiency) | All | IEEE 802.11V [3] |
| R3 (Handover efficiency), R7 (Concurrent movement) | All | IEEE 802.11R [21] |

The 802.11F WG has proposed several communications schemes between the APs involved in a particular handoff event. The 802.11 receivers often compare the Received Signal Strength against a threshold value to decide about the need to perform a handover. However, the generalized use of this signal is difficult because its range depends on each equipment vendor. It is also very difficult for a MN to detect deterioration in the communication quality, because the signal strength fluctuates abruptly due to both distance and interference. In this way, [25] proposes the number of frame retransmissions as a new decision criterion to trigger the handover in a more reliable and realistic ways, covering eventually our requirement R3 (handover efficiency).

Up to now, the most part of the handover schemes in IEEE 802.11 is mainly managed by the network side [4]. Nevertheless, the IETF Detecting Network Attachment (DNA) group [5] proposes a paradigm shift. They study how Link-layer event notifications from various wireless access technologies can allow terminals to detect as quickly as possible the eventual change of subnet. The main goal is to satisfy requirement R2 (route update). A related proposal to increase the efficiency of link-change detection is available in [38].

The wireless network management WG IEEE 802.11V supports load balancing among APs. The main goal here is to enhance the usage of available network resources (i.e. R3, handover efficiency) in a distributed way. Distinct work that also tries to fulfill the last requirement is available in [3], mainly interested in reducing the connectivity interruption on real-time traffic. Noticeably, such optimization is outside the scope of 802.11 specification referred in [16]. According to [3] the most relevant way to achieve fast handoff in 802.11 networks is to reduce the probe delay because it is the largest one. In addition, when a MN has no available access network, it can change to an Ad Hoc operation mode. In this way, it can discover a neighbor terminal that acts as a relay node with an available routing path to the network infrastructure [11].

IEEE 802.11R [21] supports connectivity aboard vehicles in motion, with fast and seamless handoffs among APs. The future network requirements supported by this amendment are R3 (handover efficiency) and R7 (concurrent movement).

Deployment Analysis

In this sub-section, we discuss potential issues that can arise on when wifi equipment is deployed at a real network infrastructure, and how wifi technology can be enhanced or modified to support some of the already evident future network requirements such as, coverage, mobility and high data rates.

A typical Wifi network infrastructure, for example, deployed at a University campus, it has a variable number of wireless controller nodes, depending on the number of expected users. However, typically there are only a few of controllers because each controller can manage up to 100 access points, using for example the IETF standard CAPWAP [62]. A controller allocates for each AP the best channel and controls the AP power transmission to minimize the Radio Frequency (RF) interference, enhancing the wireless coverage.

The AP is normally a lightweight node, without any local intelligence to manage the local network, requiring in this way the assistance of an additional controller unit, normally deployed at an Access Router of the network topology. In addition, a selected AP can be used as a dedicated spectrum analyzer in a certain area, looking for sources of RF interference, like for example wireless surveillance cameras or unauthorized APs. Then, the AP analyzer can report any problem to the associated controller. Finally, each AP can operate in terms of RF in one of two possible frequencies: 2.4 GHz or 5.4 GHz. The former offers three non-overlapped channels and the latter eleven non-overlapped channels.

There are a number of common problems that can arise when a network infrastructure uses the Wifi technology to support mobile terminals: RF interference, load demand, network performance and mobility support. These problems are following discussed in a more detailed way. Firstly, surveillance cameras and other equipment operating in the frequency range of 2.4 GHz can easily interfere with the normal operation of an AP. This interference induces transmission errors on the Link Layer, which adversely affects the performance of a TCP source, because the TCP congestion algorithm erroneously interprets the spurious link-layer errors as a network congestion problem, decreasing abruptly the TCP transmission rate and afterwards initiating a slow process to increase that rate [47]. This performance problem can be attenuated through a cross layer signaling between the Link Layer and Transport, signalizing the occurrence of link errors and consequently avoiding the unnecessary decrease on the TCP rate.

Secondly, the current popularity of new generation terminals such as smartphones could also be a threat to a Wifi network as the Wifi user admission algorithm checks the number of terminals attached to each AP against a threshold value that is configurable by the network administrator. In this way, when at a certain location there is a considerable number of smartphone users with the 2.4 GHz Wifi interface on (but not necessarily exchanging traffic with the Wifi network), these terminal interfaces become naturally connected to a local AP, which can eventually disallow other users, for example with laptops, getting a network access through that AP. One possible solution for this problem is to deploy in the neighborhood of the previous 2.4 GHz AP a redundant AP operating on the 5.4 GHz band. In this way, a laptop with a interface hardware that could switch between both RF bands could be set up to choose preferably APs working in the 5.4 GHz band, avoiding in this intelligent way the last blocking problem.

Thirdly, the Wifi backhaul access can be a network bottleneck or not depending on the entity responsible for that network. As an example, if the Wifi infrastructure covers a University campus, normally the cost of the backhaul link is not an issue here. In this way, the backhaul link is over-provisioned for the expected traffic load. In opposition, if the Wifi covers a railway station at a small/medium city, the cost of the backhaul link normally is very expensive. In this way, the backhaul link is under-provisioned and it could easily become overloaded in flash crowd situations. Further, a recent work illustrated that in Wifi networks there is an asymmetry between the uplink/downlink traffic (downlink traffic higher than uplink one), which combined with the DCF policy of providing equal opportunity for channel access to both AP and terminals results in backlogged packets at AP's transmission queue and subsequent packet losses. This results in maximum performance loss for that environment. Consequently, a solution to solve that problem was proposed that adaptively prioritizes AP's channel access over competing terminals following the downlink load [75].

Finally, the handover delay suffered by a mobile terminal could be very significant in the case that handover is between APs managed by distinct controllers. In this way, there is a considerable amount of signaling traffic between these controllers to keep the IP address of a mobile node unchanged in spite of the terminal movement. Nevertheless, the previous functionality gives a positive contribution to a macro-mobility protocol (MobileIPv4) because it avoids a new CoA registration in the Home Agent. The handover delay could be further reduced using a make-before-break handover mode. This mode does not necessarily imply that the terminal should have more than one Wifi physical interface. In fact, it is possible with a single physical wireless interface and an intermediate layer below IP to create distinct virtual wireless interfaces operating in different channels [30]. In this way, as the terminal receives beacon messages with a weak physical signal from the current AP, other APs are scanned in advance using virtual interfaces except the one currently used to exchange data with the current AP. This anticipated discovery of new APs can diminish the handover delay and support seamless handovers. Nevertheless, a real implementation of this solution is needed to clarify namely the issues of multitasking and interference.

## 3GPP LTE

The current section discusses 3rd Generation Partnership Project (3GPP) LTE, which has been recently recognized as an IMT-Advanced technology [97]. This access technology is presented, discussed, and analyzed in the next three perspectives: how the technology evolved and its expected evolution (i.e. background), how the operation of this technology can affect (positively or negatively) the players expectations discussed in the beginning of this chapter (i.e. critical analysis), and some potential issues associated to the practical deployment of a 3GPPP access technology.

<u>Background</u>

This sub-section aims to discuss in a tutorial style the most important layer 2 functional mechanisms to enable 3GPP LTE compatible equipment to be incorporated in a heterogeneous network access infrastructure, offering a broadband mobile connection.

Third-Generation (3G) wireless system, based on Wideband Code-Division Multiple Access (WCDMA) was a very popular radio access technology, and it evolved in terms of its major requirements. In this way, 3GPP has initially enhanced WCDMA to the Highspeed Packet Access (HSPA), through the introduction of High-Speed Downlink Packet Access (HSDPA) and High-Speed Uplink Packet Access (HSUPA) [54]. These technologies provided 3GPP with a radio access with higher data rates respectively in the downlink and uplink traffic. Next 3GPP Release 8 has introduced both Long Term Evolution (LTE) and System Architecture Evolution (SAE) to simplify and further improve the UMTS access. To achieve this enhancement, some actions have been taken namely, lowering costs, enabling a better integration with other standards (e.g. Wimax), and improving spectrum usage via cognitive radios [101].

Figure 3 illustrates the work that has been completed to simplify and optimize the 3GPP network infrastructure, from a hybrid and complex architecture formed by two distinct parts, UMTS circuit and packet switching network (i.e. 3GPP Release 6), to an unique all-IP flat architecture system (i.e. 3GPP Release 8 or LTE). In LTE (the right architecture in Figure 3), the Rel-6 nodes GGSN, SGSN, and RNC are assembled into a unique node, the Access Core GateWay (ACGW). In this way, the LTE architecture is simpler and flatter than the one specified in the Release 6.

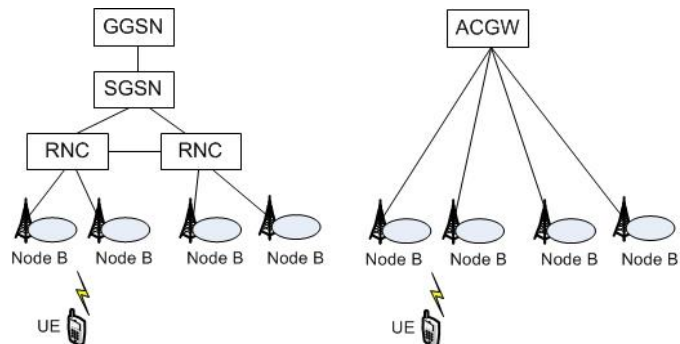

Figure 3 - 3GPP architectures: release 6 (left) and release 8 (LTE) (right)

In Figure 4, it is presented the same LTE architecture but in a more detailed way. This architecture has two distinct parts: the core one (Evolved Packet Core - EPC) and the radio one (Evolved UTRAN - E-UTRAN). As visualized in the top of Figure 4, the EPC consists of a control-plane node designated by Mobility Management Entity (MME) that manages the mobility; and two user-plane nodes both responsible for managing the exchange of data: Serving-Gateway (S-GW) and Packet-data network Gateway (P-GW). The communication between the control and user planes is supported through the interface S11. The LTE radio-access network (bottom part of Figure 4) is formed by several base stations, designated as enhanced NodeBs (eNBs), which communicate among them through the X2 interface. The eNBs are also connected to the EPC via the interface S1. In this way, an eNB can use one of four possible interfaces, depending on the information type that eNB needs to exchange (i.e. data or a control) and which node is the recipient of that information (i.e. other neighbor eNB,

MME or S-GW). In addition, each eNB device handles diverse functions such as compression of headers, security, and Automatic Repeat Request (ARQ). The mobile terminal (not shown in Figure 4) is normally denoted as User Equipment (UE). Finally, the EPC can also communicate with non-3GPP radio-access networks through interfaces S5/S8 and P-GW.

Some properties of the LTE L2 layer are now discussed in the following paragraphs. A more detailed description is available in [55]. Referring briefly the physical aspects, the LTE standard proposed the usage of Orthogonal Frequency Division Multiplex (OFDM) to modulate the physical transmission of data. In this way, LTE can: i) allocate distinct subcarriers conveniently among users according to their demands; ii) saves the energy of terminal battery; and iii) aggregates streams in favorable environmental conditions by utilizing Multiple-Input Multiple-Output (MIMO) transmissions to further increase the data rate [54].

Continuing our analysis about LTE evolution but now at layers over the physical layer, the reader could consult Table 7 where is given a summarized perspective of what in the following text will be discussed. To complement this perspective, the reader could consult 3GPP documentation about further standardization enhancements obtained by each LTE Release [108] during the last years and its future evolution.

The Release 8 introduced a significant evolution on all the three fundamental parts of a mobile network: radio access technology, core network and services. In this way, the new radio access technology of Release 8 (as already mentioned) was designated by LTE. In addition, the new core specification was entitled System Architecture Evolution (SAE), as described before. The final part of this evolution was in the services area by proposing a framework designated by IP Multimedia Subsystem (IMS).

In Release 8 it was specified the basic functionalities for the support of Home Node B (HNB) and Home eNodeB (HeNB). This was to enable the deployment of femtocells at domestic environments. The femtocells can interconnect with the 3G core and Evolved Packet Core respectively over a fixed broadband access network (e.g. DSL, Cable). Then Release 9 built on these foundations and added further functionalities that enable mobile operators to provide more advanced services as well as improving the user experience. In October 2010, Release 10 (LTE-Advanced) has been selected as an IMT-Advanced technology by ITU-R. In Release 10, it was specified a mechanism for a UE to simultaneously connect to a 3GPP access and Wireless LAN (WLAN) and transmit/receive traffic belonging to different IP flows through different wireless link accesses. The studied mechanism enables both seamless and non-seamless IP flow mobility between a 3GPP access and WLAN. Seamless offload indicates the capability to seamlessly move one or more selected IP flows from a 3GPP network to WLAN (and vice-versa) while providing IP session continuity on each flow. This seamless offload was based on DS-MIPv6. Non-seamless offload indicates the possibility to exchange the traffic of one or more selected IP flows using WLAN IP address (referred also as Direct IP Access) without providing any IP session continuity. Further discussion on mobility support is available in [97].

In Release 11 was studied the provision of machine-type communication services at a low cost level through mobile networks, to match the expectations of mass-market machine-type services and applications. An example of this service is related with consumer products manufacturers that aim to be in touch with their products after they are shipped – car manufacturers. Another example is in the home environment where remote maintenance of heating and air condition, alarm systems and other applications can also be identified.

In Release 2012 was concluded the study initiated in Release 11 about roaming end-to-end scenarios with VoLTE IMS and other networks with local breakout, which uses a device designated by Breakout Gateway Control Function (BGCF). The BGCF is a SIP proxy which processes requests for routing from an S-CSCF when the S-CSCF has determined that the session cannot be routed using DNS. The S-CSCF is the proxy server controlling the communication session inside a specific domain.

In Release 13 is being studied the management aspects of Network Functions Virtualization (NFV) in 3GPP. The expected advantages of adopting NFV in mobile networks are the following ones: network functions can be easily scaled in and out dynamically; reduce the time for deployment of new services; the decoupling of hardware and software allows the reduction of space, power and cooling requirements and hardware diversity. In addition, this Release is studying how to support machine-type communications with low throughput (i.e. 160bps), low complexity, in a scalable way with scenarios with two persons at each home and each person is using 20 Internet of Things (IoT) devices, and cutting down the power usage of devices (e.g. by optimizing signaling exchanges in the system to realize battery life of up to ten years).

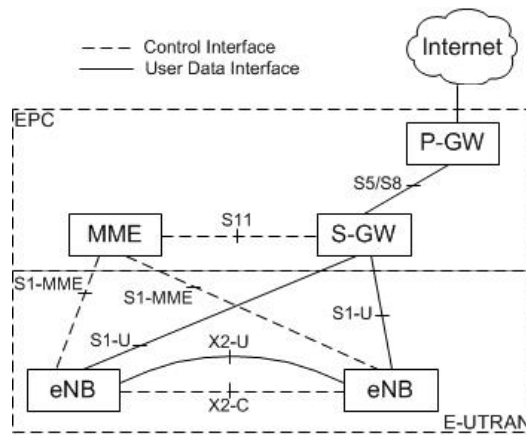

Figure 4 - LTE architecture in more detail

Table 7 – 3GPP LTE Evolution

| 3GPP LTE Release | Main Goals |
|---|---|
| 8 (2008) | Introduced an evolution on all the three fundamental parts of a mobile network: radio access technology, core network and services. It was also specified the basic functionalities for the support of Home Node B (HNB) and Home eNodeB (HeNB). |
| 9 (2009) | It was added to Home Node B (HNB) and Home eNodeB (HeNB) further functionalities such as more advanced services and for improving the user experience |
| 10 (2011) | Specified a mechanism for a UE to simultaneously connect to a 3GPP access and WLAN and transmit/receive traffic belonging to different IP flows through different accesses. |
| 11 (2012) | Provision of machine-type communication services through mobile networks. |
| 12 (2014) | Conclusion of the study on Technical aspects on Roaming End-to-end scenarios with VoLTE IMS and other networks. |
| 13 (Late 2015) | Study on Network Management of Virtualized Networks; Study on Cellular system support for ultra Low Complexity and low throughput IoTs |

Critical View

In this sub-section, it is discussed how the LTE architecture impacts some of the future network requirements already discussed. Some main goals of the LTE architecture are reduced latency and cost [24] as well as enhanced data rate, larger capacity and spectrum efficiency. Based on these, it is perfectly reasonable to consider a system architecture with a low number of nodes. This reduces the complexity and cost for testing a network infrastructure based on LTE technology. In addition, the latency of traffic traversing a LTE infrastructure can be reduced. This latency reduction has a positive influence in terms of requirements R2 (route update) and R3 (handover efficiency) because the new path to a mobile node could be updated more quickly during a handover (see Table 8).

As there are fewer nodes in the LTE architecture, LTE could be less robust against a node failure or a link congestion (negatively impacting R6) and it could have some scalability problems with a large number of simultaneous handover requests (negatively impacting R9). Nevertheless, these problems can be attenuated because the LTE architecture has independent network entities to manage the mobility and the traffic data. In addition, as a more flexible hierarchical ARQ functionality is available at layers RLC and MAC, the requirement R6 (robustness) could be positively impacted because the most part of the link transmission errors is avoided by the lightweight HARQ protocol of MAC layer. The last term, lightweight, means that the protocol HARQ does not overload so much the network like ARQ does at the RLC layer and HARQ detects more quickly eventual errors.

In terms of the equipment to deploy LTE, this architecture requires new self-adaptive radio equipment to maximize the spectrum usage  as well as new equipment in the network part of the infrastructure. Consequently, the requirement R8 (deployment factors) is impacted negatively because operator cost can be significantly increased.

As the link cost between each eNB and S-GW is very high, these links are unlikely to be over-dimensioned. This fact implies these links can become congested very often, originating packet loss that negatively affects R1 (message forwarding).

The movement of handsets within the area covered by LTE radio equipment is controlled by Link layer mechanisms and a direct communications interface among eNBs. This affects positively the requirement R3 (handover efficiency) as handover delay and packet loss can both be effectively reduced.

The eNB performs encryption and integrity services on the control and user data planes. This enables a solution with a security level similar to the legacy cellular architecture, giving a positive contribution on requirement R5 (security).

Deployment Analysis

In the current sub-section, we discuss the deployment of 3GPP access technologies in the following aspects: interoperability, mobility support, and network coverage. In the first aspect of interoperability, 3GPP a long time ago has been interested to support the roaming of users between 3GPP networks and non-3GPP networks. In 2005, 3GPP has incorporated in its standard the Unlicensed Mobile Access (UMA) concept that provides user roaming between GSM/UMTS, WLAN, and Bluetooth networks. In parallel, there are several WLAN/3G integration architectures proposed in the research literature, based on the inter-dependence between WLANs and 3G networks [6].

Secondly, the mobility of cellular hosts within 3GPP GPRS and UMTS release 5 networks is handled by link layer mechanisms [26]. Considering this and the interoperability aspect, there is a need to develop a cross-layer design among the link layer and upper layers to support efficient handovers managed with a macro-mobility protocol, e.g., MIPv6.

Thirdly, other access technologies can also complement the coverage offered by a 3GPP network in various scenarios, as for example, mesh networks [48], femtocells [21], an optical fiber [53] or hotspots [47].

Table 8 - LTE impact on future network requirements

| Functionality | Affected Requirement | Affected Player | Impact Description |
|---|---|---|---|
| Architecture with less nodes | R2+, R3+ | All | A new routing path to UE is updated quickly; Minimizes packet loss and delay during handover |
| Architecture with less nodes | R6- | Network Provider | If P-GW fails then LTE have no communications with other networks |
| Hierarchical ARQ available at layers RLC and MAC | R6+ | Network Provider | Most errors are solved by lightweight HARQ protocol of MAC. Only residual HARQ errors are managed by RLC |
| LTE architecture is more flat | R9- | Network Provider | Scalability problem with a high number of handovers |
| New self-adaptive radios are needed to maximize the spectrum efficiency | R8- | Network Provider, Terminal | It increases operator cost, terminal cost |
| The link cost between each Node B and S-GW is expensive; it is expected the under-provisioning of these links | R1- | All | Packet losses will occur when the links become congested |
| The movement of handsets within the network is controlled by link-layer mechanisms and eNBs interface | R3+ | All | Handover delay and packet losses are reduced |
| eNB performs ciphering and integrity protection of control and user data planes | R5+ | User | It enables a security solution at least as strong as legacy architectures |

## IEEE 802.16 (WIMAX)

The current section discusses IEEE 802.16, which has been recently recognized as an IMT-Advanced technology [97]. This access technology is presented, discussed, and analyzed in the next three perspectives: how the technology evolved and its expected evolution (i.e. background), how the operation of this technology can affect (positively or negatively) the players expectations discussed in the beginning of this chapter (i.e. critical analysis), and some potential issues associated to the practical deployment of Wimax.

Background

The main goal of this sub-section is to discuss in a tutorial style the most important layer 2 functional mechanisms to capacitate a wireless network infrastructure based on IEEE 802.16 access technology to offer a broadband mobile connection in cooperation with other access technologies.

IEEE 802.16 (or Wimax) technology offers broadband connectivity to both fixed users (IEEE 802.16) and mobile users (IEEE 802.16e or Wimax 1.0 – see Table 9). The 802.16 supports both a connection-oriented MAC and a differentiation between the downstream and the upstream traffic exchanged between a Base Station (BS) and each terminal, via a single-hop wireless channel. The Wimax uses a QoS mechanism based on distinct downstream connections between the BS and the user terminal. Each connection could use a specific scheduling algorithm according the quality requirements of the data traffic to exchange via that connection. In parallel, each terminal can request from the serving BS additional upstream channel bandwidth. There are two modes that a terminal can choose to transmit a bandwidth request to the serving BS: contention mode (best effort) and contention-free mode (polling). The contention-free mode is more convenient for QoS-sensitive applications due to offering a more predictable delay. Finally, IEEE 802.16e enables mobile stations to handover among base stations in two modes: hard and soft. The last mode is recommended to satisfy the more restrictive requirements of VoIP applications.

The next release was Wimax 1.5 concerned with the loosely coupled method of interworking between WiMAX systems and cdma2000 systems. This architecture is applicable to an operator that owns both access technologies and provisions its users with a dual mode device (dual radios) that can connect to the core network through any one of the two technologies. In this way, the handovers among technologies use a make before break methodology to support a seamless session continuity. Essentially, this means keeping the IP address (Home Address in Mobile IP) assigned to a MS at one point of attachment so that it can continue to send and receive packets for an ongoing IP session, in spite of the user terminal handover between technologies. In addition, the seamless handover means to minimize packet loss during a change in point of attachment during a handover process. To perform a seamless handover both interfaces of the MS should be maintained active for a period of time when adequate and overlapping coverage is available between old and new attachment points. Thus, Mobile IP4 or other registration procedures can take place on the new radio interface while packets are still being sent and received on the old radio interface. In order to minimize packet loss during inter-technology handover, the HA can accept packets from the previous Care-of Address for a limited period of time.

Table 9 – Wimax Evolution

| Release | Main Goals |
|---|---|
| 1.0 (2004) | Mobility is supported as well as some enhancements on downstream/upstream connections with some guarantees concerning QoS |
| 1.5 (2009) | Loosely coupled method of interworking between WiMAX systems and cdma2000 systems using simultaneously two radios |
| 1.6 (2011) | Single radio handovers among distinct access technologies in multi-mode terminals |
| 2.0 (2012) | Mobility management in dual-radio mode is supported in both versions of IP. |

A multihop relay amendment was approved in 2009, IEEE 802.16j. In this scenario, mobile stations can act as relays, forming a multi-hop network between mobile terminals localized at the cell's edge and the serving BS of the relay participating in the last (upstream) on-the-air hop. In addition, BSs can communicate via a backbone that can be either wired or wirelessly.

The Wimax 1.6 release was concerned with the fact that there is a strong need to support Single Radio Handovers between Wimax and Non-Wimax Access Networks in multi-mode devices for the following reasons: only one radio can operate satisfactorily at any given point of time due to co-existence, interference, noise and other such issues for radios operating in close frequency ranges; to increase the battery autonomy of multi-mode device; due to regulatory and other issues, simultaneous multi-radio operation may not be always possible.

To support single radio handovers from Non-Wimax IP Access Network to Wimax was proposed a new functional element designated by Signal Forwarding Function (SFF). In this way, SFF facilitates pre-registration and authentication while the UE/MS is connected via the non-Wimax Access Network prior to active handover to the Wimax Network. Further information is available in [109].

The 802.16m amendment (Wimax 2.0) enables more efficient and fast data communications. The equipment manufactured accordingly the more recent standard should also support legacy equipment.

Wimax 2.0 specification (available from [109]) assumes that the mobile terminal only operates in dual-radio mode i.e. both radios can transmit and receive simultaneously. This specification also assumes that a dual mode mobile terminal is connected to a common 3GPP Core (EPC) via Wimax ASN. Scenario where dual mode mobile terminal is connected to a common Wimax Core (CSN) via 3GPP access is not supported. The mobility management is supported in both IPv4 and IPv6 respectively using MIPv4 (CMIPv4) and Proxy Mobile IPv6 (PMIPv6). The 802.16m (WirelessMAN-Advanced Air Interface) has been selected as an IMT-Advanced technology by ITU-R, in October 2010.

Critical View

In this sub-section, it is discussed how the IEEE 802.16 (Wimax) architecture impacts some of the future network requirements already discussed. Wimax uses at link layer both ARQ and HARQ protocols in a similar way as already explained for LTE. The previous functional characteristics of Wimax have a strong impact on requirement R9 (scalability) because a single BS can manage a large number of SSs, as shown in Table 10.

The IEEE 802.16e-2005 supports mobile users and is often referred to as mobile Wimax. In addition to the fixed broadband access, mobile Wimax proposes four typical mobility scenarios [28] [102]: i) Nomadic mobility where the user terminal manually connects to the more convenient network point of attachment; ii) Portable mobility where a best-effort network access is provided to a terminal moving at a speed lower than 3km/h); iii) Simple mobility where the subscriber terminal may move up to 60km/h with brief interruptions (less than 1sec) during handover; and iv) Full mobility where the subscriber terminal may move up to 120km/h and seamless handovers (less than 50ms latency and <1% packet loss) are supported.

As Wimax supports distinct mobility usage scenarios it can have a positive impact on several future network requirements. The first requirement that can receive a positive contribution is R2 (route update) if the mobility agents are updated with a new route to the mobile node according to the handover requirements of each one of the mobility scenarios listed above. In the case of a seamless handover, a large number of successful packets are delivered to a mobile SS, giving also a positive contribution to requirement R1 (message forwarding). The requirement R3 (handover efficiency) has also a positive contribution because both the handover delay and lost packets are minimized.

Table 10 - Wimax impact on future network requirements

| Functionality | Affected Requirement | Affected Player | Impact Description |
|---|---|---|---|
| Differentiates upstream and downstream + Connection oriented + MAC uses ARQ and HARQ | R9+ | Network Provider | A large number of SSs served by a single BS |
| Supports several mobility scenarios | R2+ | All | The route updates can be prioritized according the handover requirements of each scenario |
| Supports several mobility scenarios | R1+ | All | Seamless handover delivers a large number of successful packets to SS |
| Supports several mobility scenarios | R3+ | All | Minimizes packet loss and handover delay |
| Signaling mechanisms to track SSs | R4+ | Terminal | SS battery can be saved as SS could be easily followed |
| Supports MIP | R5- | User | Exposure of MN location to CN |
| Supports MIP | R7- | All | Concurrent movement of both SS and CN difficult to support |
| Supports MIP | R9- | Network Provider | Scalability problem |
| Architecture proposes the same network entity to manage mobility and data | R9- | Network Provider | Scalability problem with a large number of handover requests simultaneously with a high network load |
| Architecture proposes the same network entity to manage mobility and data | R6- | Network Provider | If ASN fails then Wimax cannot operate well and it can't communicate with other networks. |
| Wider frequency range | R8- | Network Provider, Terminal | Requires new and expensive equipment |

The Wimax defines signaling mechanisms for tracking subscriber terminals as these move from the coverage range of a BS to another, including in the extreme case when a terminal is temporarily idle to save the energy of battery. The previous signaling mechanisms to track the SSs can have a positive contribution to requirement R4 (mobile node location – battery autonomy).

The Wimax architecture supports end-to-end IP-layer mobility using mobile IP [28]. Consequently, 802.16e can inherit some drawbacks from MIP that are discussed with more detail in a following section dedicated to MIP: exposure of the MN location to the CN (R5 -, negative impact on security), support of simultaneous movement of MN and CN (R7 -, negative impact on concurrent movement) and scalability problem (R9 -, negative impact on scalability).

The Wimax architecture proposes a single network entity (i.e. ASN) to manage both mobility and data traffic. This architectural choice can lead to a scalability problem when a large number of handover requests occur simultaneously with a high network load (R9 -, negative impact on scalability aspects). In addition, if the previous management network entity fails then the Wimax network infrastructure can be adversely affected by that and it has no communications with other networks. To avoid these last problems a redundant ASN node is required in the network infrastructure.

The Wimax standard proposes a wider RF range of than the one proposed for Wi-Fi compatible equipment (i.e. typically 2.4 GHz). This will drive to the need for a highly complex and cost effective technological solution inside Wimax/Wifi compatible handsets (R8 -, negative impact on deployment factors). Nevertheless, this tremendously technological challenge is at the time of the current writing being successfully solved by the hardware manufacturers [96]. Further, the Wimax standard offers an interesting feature to the network providers, whom decide to upgrade their network infrastructures, to support in the same network infrastructure any combination of new and legacy node devices [93]. In this way, the network providers can perform a smoother financial effort to update their networks.

<u>Deployment Analysis</u>

In this sub-section are highlighted some open research issues to enhance the deployment of Wimax. Firstly, it is important to study efficient mobility solutions for Wimax topologies different from the point-to-multipoint one, like the mesh wireless topology. The main goal is to enable Wimax to cover efficiently the last mile connectivity problem [8] [34] because the LTE deployment could be very expensive.

Secondly, there is also the work of the IEEE 802.20 Mobile Broadband Wireless Access (MBWA) standard, which is very ambitious and wide-ranging in scope [7]. This standard is trying to be the best of all worlds — providing users with a high bandwidth, low latency, and always-on Internet service at home and whether they are commuting or at work. This technology could fill the gap between cellular and WMAN/WLAN services and solve the problem of widespread mobile Internet usage. Nevertheless, only the future will tell whether MBWA will prove to be feasible from both technological and engineering standpoints, as well as from the standpoint of being a viable business model. For example, the authors of [7] have the opinion that 802.20 will have to compete with home networks that use traditional cable and DSL broadband technologies.

Finally, the communications national regulators are opening the spectrum to be used by any type of technology but in a controlled way. As an example, Ofcom (UK) launched a public consultation on the award of the band 2500-2690 MHz [35]. Ofcom proposed the spectrum was awarded in a technology neutral way. A large diversity of technologies is planning to use this spectrum like 3G or its evolutions and wireless broadband services such as Wimax, IEEE 802.20 or ETSI HIPERMAN. Following the same efficiency strategy to reuse available spectrum in bands not initially selected for a specific access technology, we can also pointed out the case of Wifi where is going on an IEEE standardization work to produce a new amendment (i.e. ah) to enable a Wifi network operating in sub 1 GHz license-exempt bands.

## WIMAX and LTE Developments to Support Mobility

Both Wimax and LTE have recently being considered by International Telecommunication Union - Radiocommunication sector (ITU-R) as International Mobile Telecommunications (IMT) – Advanced systems [60]. This section compares recent developments in both Wimax and LTE related with the mobility support. In addition, it is discussed how the handovers between these two technologies can be supported.

The authors of [97][93] explain and compare the state-of-the-art handover schemes developed for LTE and Wimax networks. In IEEE 802.16-2009, there are described distinct deployment ways to optimize the handovers [97]. At one end, no optimization is used, and in such a configuration a mobile node has always to perform a full network entry in the new BS (with a consequent degradation on handover latency) regardless of maintaining/updating its IP address. At the other end, there is the "fully optimized handover" where the MN's context information during the handover is moved between the serving BS and the target BS, and transparently to the mobile node. In this way, the handover latency is reduced as the mobile node does not need to make a full network entry. The same standard defines two more (soft) handover modes [97]: Macro-Diversity Handover (MDHO) and Fast BS Switching (FBSS). Both of these specify a list of BSs with which a mobile node can connect to. The difference between them is that in MDHO the mobile node communicates with all BSs of that list, whereas in FBSS the mobile node only communicates with the anchor BS [97]. An important advantage of FBSS is that if the target BS is within the BS list then the mobile node does not need to perform a full handover procedure because it merely changes the anchor BS.

The IEEE 802.16j amendment describes how the multihop relay BSs and relay nodes should operate during the movement of mobile nodes, and moving relay nodes. In both cases, the multihop relay BS maintains substantial control of handovers [97].

The IEEE 802.16m amendment endorses different mobility scenarios with only 16m BSs (single or multicarrier) or with a mixture of 16m and legacy BSs. In addition, the 802.16m discusses the possibility of enhanced inter-RAT handover procedures when the mobile nodes have a single or multiple interfaces. Further, this amendment discusses several mobile scenarios with femtocells [97].

A novel functionality introduced by 802.16m enables BSs and terminals to use multiple carriers (below the MAC layer) for high capacity-connectivity inside a cell. Consequently, terminals with multicarrier capability can perform an entry-before-break handover by maintaining a connection (using a specific carrier) with the current BS and performing in parallel and advance a network re-entry at the target BS, using a distinct carrier. This eliminates both the handover delay and data loss during the handover process. The release 10 (i.e. LTE-Advanced) is studying carrier management during the handover in a similar way to the one previous explained of 802.16m [93].

The authors of [93] argue that LTE supports two more handover modes: seamless and lossless handover. In this way, the LTE seamless handover mode can be used by VoIP, which is a delay-sensitive traffic but tolerant to packet losses. In contrast, the lossless handover can be used with delay-tolerant applications, such as File Transfer Protocol.

The handover between 802.16m and LTE is performed very distinctively depending on the technology which receives a new mobile node (from the other). This occurs because both organizations (IEEE and 3GPP) are working on standards which are not fully compatible, as we following detailed [93]. On one hand, to manage a handover from 802.16m to LTE, two

L2 protocols are proposed by IEEE. The first transfer protocol uses an 802.16m generic MAC L2 transfer tunnel that allows transfer of signaling traffic between a mobile terminal and the CNs directly for handover initiation and execution. The second protocol uses the 802.21 standard [9] to allow communication between the mobile terminal and an 802.21 server that acts as a proxy for the terminal to prepare the initiation of the handover with other involved network entities. One should note that the 802.21 only supports handover initiation; it does not perform the handover. Therefore, to perform the handover itself an adequate mobility protocol should be operating together with 802.21.

On the other hand, for a handover from LTE to 802.16m, only a handover via L3 is currently available because 3GPP does not specify any L2 radio access signaling to the serving (LTE) eNodeB could interact with the target (Wimax) BS. By other words, from E-UTRAN's network point of view, any handover that has its final destination a node belonging to a Wimax infrastructure is completely transparent to the (LTE) eNodeB. As a conclusion, assuming the fact that a vertical handover between the two access technologies discussed in the current section is managed not only by different ways but sometimes also using different layers, the vertical handover delay could considerably vary, depending on the target technology.

Further details about the interworking among heterogeneous access technologies can be found in [77].

## Some Thoughts about the Evolution of Wireless Access Technologies

The authors of the current chapter think that the ubiquitous and mobile access to a NGN environment should gradually be supported by a set of distinct wireless technologies fulfilling 4G (and beyond) requisites. In this way, all these 4G access technologies will share their available connectivity resources, enabling the mobile terminals to use efficiently the entire available network connection capacity of the heterogeneous infrastructure.

Emergent virtual and/or social communities will share a large amount of data and services, including multimedia. In this way, it is very important to develop adaptive multimedia management solutions for ensuring the best possible quality to the end-user flows according to the complete set of available network resources across the heterogeneous network infrastructure, including low-range communications among the mobile terminals.

To enable upcoming 4G (and beyond) requirements, the connection service should become gradually more symmetric (i.e. uplink and downlink with similar throughput). This will also require a more efficient management of terminal battery, eventually exploiting the cooperation among mobile devices. Some 4G research and standardization work is discussed in more detail in [44].

In the highly dynamic world of the NGN environment, any terminal could transmit, in any particular moment of time, choosing the most adequate transmission technology among several or, using multiple paths in parallel through distinct technologies to increase the throughput [74][76]. In addition, an efficient load balancing mechanism among heterogeneous wireless networks is required to enable the deployment of emergent mobile applications [2]. The general expectation, therefore, is that of co-existence, where each 4G (and beyond) wireless access technology is deployed to fulfill application requisites but in a compatible (and complementary) way with other 4G access technologies.

# MOBILITY MANAGEMENT AT THE NETWORK LAYER

As the Internet Protocol (IP) provides internetworking in the Internet, solutions deployed at the network layer are classified a good option to support mobility. In addition, a mobility protocol implemented on top of the link layer also becomes naturally independent of any new (version/release of) wireless access technology. We next discuss Mobile IPv4, Mobile IPv6 and various prominent extensions to these protocols. For each selected solution, we present that solution initially in a tutorial manner, then we criticize the same solution, highlighting its strengths and weaknesses according respectively how well that solution addresses some expected relevant requirements for future networking environments. Finally, we analyze in a single section the deployment of all the selected mobile proposals at the network layer and compare them.

## MIPv4

IETF MIPv4 was the first robust proposal to support mobility among distinct network domains, using the network layer to accomplish that.

<u>Background</u>

The main goal of this sub-section is to discuss in a tutorial style the most important mechanisms of MIPv4 to support mobility.

In order to support MIPv4 (Figure 5) each network should have two agents with different roles: Home Agent (HA) and Foreign Agent (FA). The HA in the Home domain is responsible to capture any packet destined to a local IP address of a terminal (currently attached to a different domain) and sent it to the network where that terminal is connected to. In this last network, there is a FA that discovers new visiting terminals and then can perform their registration into their HAs. In this way, each mobile terminal can always use the same Home IP address independently of the network is attached to.

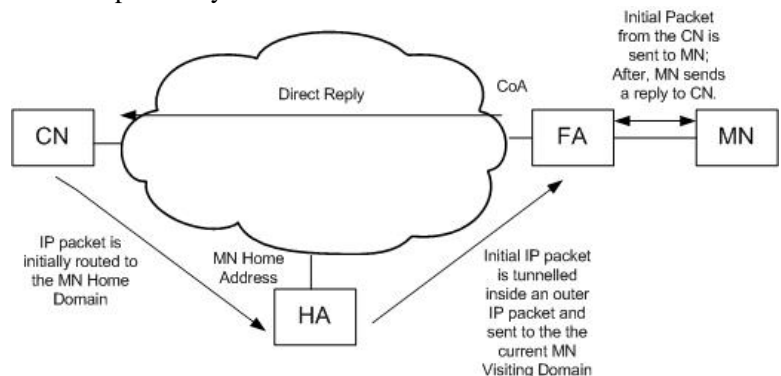

Figure 5 - MIPv4 Architecture

In MIPv4, as the MN is connected to its home network then no special support from MIPv4 is necessary. If the MN is out of its home network and MN connects to a new network, then the MN obtains from the FA a Care of Address (CoA) through an ICMP route advertisement. Then, a Registration message is sent by the MN to the HA using FA as an intermediary node. This message allows the MN to update the HA with the new CoA. After the HA received this message, it creates a local entry, mapping the MN IP Home address with the CoA. To conclude this registration process (i.e. binding), it is sent a confirmation (reply) message from the HA to the FA. Then, the FA resends this last message to the MN, concluding in this way the registration process.

After the registration has been successfully executed, the HA can use the Proxy Address Resolution Protocol (Proxy ARP) to intercept packets destined to the MN that arrive to the home network. After this, the HA forwards these captured packets to the FA using a tunnel. When the FA receives the tunneled packet, it takes out the outer IP header and delivers the inner packet to the MN. When the MN sends a reply to the CN, that packet is only directly routed from the MN to the destination (CN) if the ingress-filtering is inactive in the FA router. Otherwise, as the ingress-filtering is active in the FA, any packet with an incorrect source address is immediately detected and dropped because it is erroneously interpreted as belonging to a spoofing attack. To avoid this erroneous detection, the FA, in the case the ingress-filtering is on, should not send the packet received from the MN directly to CN but tunnels it and sends the tunneled packet to the HA. After the HA just received the tunneled packet from the FA, the HA de-tunnels the inner packet and afterwards delivers this last packet to the CN. The drawbacks of this last case, in comparison with the case of a direct communication between the MN and CN, are the slightly overhead in terms of header size and a latency increase on the final deliver of the reply packet to CN.

The work presented in [11] extends MIP so that the HA as well as the CNs can maintain multiple bindings for a MN simultaneously connected to several foreign networks. The MN's home address will then be bound to multiple CoAs, each one representing a distinct domain or foreign network. To communicate with the MN, the HA as well as the CNs can then select an address among the registered CoAs. The selection of CoA is based on RTT measurements that are carried out on the MIP messages. In this way, no explicit ICMP messages have to be sent. Then, this proposal enhances network connectivity by enabling the MN, the HA and CNs to evaluate and select the best connection. It also balances the load generated by MNs between several FAs. Finally, this proposal is also resilient to either FA failure or abrupt connection lost. This work could be enhanced with a mechanism to avoid the HA failure problem, using some kind of HA redundancy. In addition, research work using other wireless access technologies besides the IEEE 802.11 is also needed (i.e. using heterogeneous mobility scenarios).

Critical View

This sub-section discusses how MIPv4 supports mobility and how it impacts some of the future network requirements already discussed. The aspects discussed in the next paragraphs are summarized in Table 11. MIPv4 has some drawbacks. Firstly, packets destined to a visiting MN are always routed via its HA, which could originate a problem normally designated by triangular routing. Alternatively, to avoid the ingress filtering executed by a firewall at the edge router of a foreign network the packets destined to CN must be initially tunneled by the FA, sent to the HA at the home network and only afterwards the de-encapsulated packets are finally delivered to CN. This affects negatively the requirement R3 (Handover efficiency), because the packets are delayed. MIPv4 is also vulnerable to the problem related with the single point of failure (i.e. HA), which affects negatively the requirement R6 (Robustness).

Secondly, MIPv4 CoA update at the HA (i.e. registration) is only performed at the time the MN visits a new foreign network. Before this visit, the HA could has been sending packets to the available CoA but this is not anymore valid because the MN was disconnected during the period of time the MN was moving between networks. This last situation causes packet loss from the traffic originated at CN and destined to the MN. In addition, the MIPv4 registration is made in a bidirectional interaction between the HA and the FA. If the routing paths between the two agents have a significant RTT then the MIPv4 registration could take a long time, which could increase the handover latency and packet loss. Furthermore, the Agent Advertisement and Registration messages overload the network infrastructure. All the MIPv4 aspects discussed along the current paragraph  have a negative impact in the requirement R3 (Handover efficiency).

If the number of mobile users raises up, then the overhead due to signaling also increases, which impacts negatively the requirement R9 (Scalability). To mitigate this, a few number of techniques, such as route optimization and, both hierarchical and micro-based mobility protocols (e.g. HAWAII, Cellular IP, HMIPv6), have been proposed to enhance the performance of MIPv4 [59]. A further discussion about distinct micro-mobility protocols can be found in [63]. An innovative study about how to optimize Mobile IPv4 handover performance for a cellular protocol is available in [23], where the overhead could be further diminished by the fusion of quasi-simultaneous mobility registration messages when they share either home and visited domains or at least the visited domain.

Table 11 – MIPv4 impact on future network requirements

| Functionality | Affected Requirement | Affected Player | Impact Description |
|---|---|---|---|
| Packet tunneling at FA | R3- | All | Increases the packet delay |
| HA failure | R6- | Network Provider | Proposal doesn't work |
| Binding update delay | R3- | All | Packet loss |
| High RTT of path between mobility agents | R3- | All | Increases handover latency and packet loss |
| High number of MNs | R9- | Network Provider | Increases signaling overhead |
| Packet reverse tunneling | R5+ | User | MN privacy is supported |
| Registration messages aren't authenticated | R5- | User | Bogus FA could stop or capture traffic |

Alongside with previous MIPv4 limitations, there are also some security problems that have a negative impact on requirement R5 (Security analysis). These problems are router's ingress filtering at the visited network, MN's authentication, and replay attacks. Considering the first security problem, as already discussed, this problem can be solved by using reverse tunneling between the FA and the HA, which increases the packet delay. Nevertheless, it ensures the positive aspect that the MN's privacy is supported because the MN location is not exposed to CN. In fact, the packets sent by MN from the foreign network and received by CN could have always the MN home IP as the source address. The second problem could occur if a bogus CoA registers with the HA. In this situation, the bogus CoA could block all the data traffic destined to the MN, or even worse, cause all packets to be redirected to a potential attacker. To avoid this, all the messages used by the registration process should be authenticated. The third problem involves an attacker resending old registration messages. To counteract this situation, MIPv4 proposes the message sender to add some extra header fields (e.g. timestamps combined with nonces) to the registration messages in order the receiver could validate each received message, detecting in this way old retransmitted messages.

## MIPv6

Some evolution has occurred from MIPv4 to MIPv6, from which one of the most notable aspects is the one related with route optimization in specific mobile scenarios.

<u>Background</u>

This sub-section discusses in a tutorial perspective the most important mechanisms of MIPv6 to support mobility.

IPv6 also supports Internet mobility management using MIPv6 [12]. Figure 6 shows the MIPv6 architecture and its main operations. MIPv6 follows the same basic principles as MIPv4, including the idea of a home address, CoA, HA, and tunneling. The main difference is that MIPv6 does not need a FA because IPv6 has new embedded mechanisms, which did not exist in IPv4: Stateless Address Auto-configuration, Neighbor Discovery and Duplicate Address Detection. In this way, a mobile node in MIPv6 can configure and verify its own CoA by using previous IPv6 native mechanisms. In addition, the mobile node can dynamically discover the home agent by adding a well-known anycast interface identifier to its home link's prefix. In the case the MN is connected to an IPv4 foreign network, it cannot use the anycast identifier. In this scenario, the MN uses DNS to discover the home agent's IPv4 address [94].

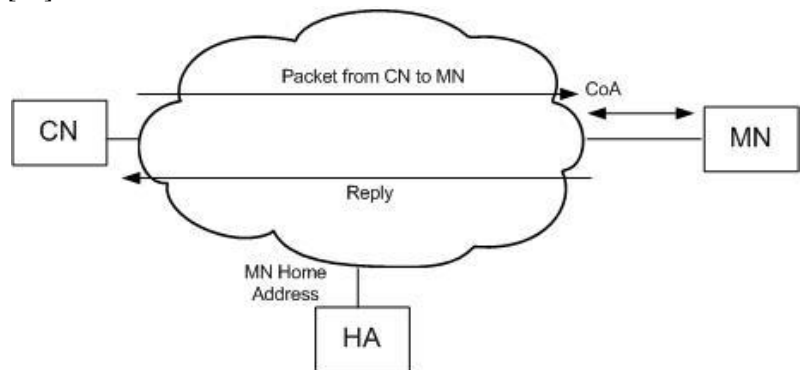

Figure 6 - MIPv6 Perspective

When the CN communicates with the MN for the first time, the first packet is routed to the home network and tunneled through the HA like already explained in MIPv4. The difference now is that IPv6 encapsulation is being used instead of IPv4 one as before (i.e. we also assume the MN has already registrated itself at the HA). After the first exchange of packets between CN and MN has been made through a tunneling technique, the MN registers to CN with a BU message. The CN stores this binding information in the binding table. After this, the MIPv6 can perform route optimization between MN and CN [92], as we following explain.

After the MN registrated itself into CN, the CN can route packets directly to the CoA of the MN with the MN's home address in the Home Address Destination option. Similarly, the MN can send packets directly to the CN using the Home Address Source option. It holds the MN's home address, so that the sending MN can insert the topologically correct CoA in the source address of the IP header to avoid the ingress-filtering problem. At the destination, the CN replaces the CoA by the MN's home address before delivering the data to the transport layer. Thus, after the last processing, the higher layers see a normal IP packet that looks like it was sent from the MN's Home Address. Further information to enhance MIPV6 route optimization can be found in [43].

Some research work has been performed to enhance MIPv6 [36] [37] [27]. The authors of [36] have proposed a scheme based on DHCP- to dynamically discover the MIPv6 home network information. Other contribution from IETF [37] describes mechanisms for home agent reliability, state transfer, and failure detection. There is also an extension to MIPv6 (MCoA) [27] that enables the registration of multiple CoAs to the same home address [27]. This could allow a multi-RAT scenario in the visited domain.

Critical View

In this sub-section is discussed how MIPv6 can impact on some of the future network requirements already discussed. The aspects discussed in the next paragraphs related with MIPv6 [12] are summarized in Table 12.

Table 12 – MIPv6 impact on future network requirements

| Functionality | Affected Requirement | Affected Player | Impact Description |
|---|---|---|---|
| Route optimization | R5- | User | MN location is exposed to CN |
| IPsec | R5+ | User | Signaling messages are authenticated, encrypted and can't be modified |
| High number of MNs | R9- | Network Provider | Increases signaling overhead |
| High number of MNs | R3- | All | Increases packet loss and handover delay |
| CN and MNs communicate directly | R9- | Network Provider | Mapping table CoA-MN Home Address at CN could be very large |
| It supports IPv4/IPv6 migration | R8+ | Network Provider, Terminal | It can smooth the investment effort needed to upgrade the network/terminal |

As it was already mentioned, there are some differences between MIPv4 and MIPv6. For example, route optimization has been incorporated into MIPv6 [92] as an alternative to the reverse tunneling of MIPv4. The main advantages offered by route optimization are lower handoff delays, increased security, and reduced signaling overhead. In this way, the MN uses its CoA as the source address in the IP header, so it can avoid both problems of triangle routing and ingress filtering, sending the packets directly to the CN through the firewall of the visited network. Nevertheless, this optimization compromises MN location privacy by potentially exposing the CoA, and hence its location, to the CN. This implies a negative impact on requirement R5 (Security analysis). This security problem could be partially solved using HMIPv6 (see above), because only the RCoA is sent in the Binding Update (BU) messages from the MN to CN. In this way, only the MN domain location is exposed to CN (and not the MN visiting network). MIPv6 also offers advantages in terms of security support when compared with MIPv4, as BUs and Binding Acknowledgment messages (BAs) could be both authenticated using one of two possible protocols proposed in IPsec, Authentication Header (AH) [45] and Encapsulated Security Payload (ESP) [46]. Both of these security IP protocols provide data origin authentication and integrity for datagrams. ESP also supports encryption at network layer. All these security aspects are strong points for MIPv6 in terms of the requirement R5 (Security analysis).

In spite of all these positive points, MIPv6 could present the drawback of significant extra signaling overhead, especially when MNs move quickly or the number of MNs significantly increases, because BUs are transferred among the MN, the HA, and the CN. Then, it could also suffer from high packet loss (due to wireless channel errors) and high handover latency problems, thereby deteriorating the user-perceived quality of real-time traffic. All these aspects impact negatively the requirement R3 (Handover efficiency), as well as requirement R9 (Scalability aspect).

MIPv6 itself is also unrealistic for service providers. For example, it seems difficult for Google servers (i.e. in the role of CNs), with millions of hits per minute being required, to support MIPv6 route optimization because the mapping table COA – MN can become extremely large. This aspect impacts negatively the requirement R9 (Scalability aspect). Other negative aspect is the loss of privacy of the MN (R5-). For solving the last scalability problem, the use of a multicast mobility proposal could be envisioned to disseminate the contents from the service provider to MNs using a few of multicast group addresses [20].

Using the MIPv6 extension, Dual Stack MIPv6 (DSMIPv6) [94], MNs (IPv4 or IPv6 aware) would only need MIPv6 (and NEMO for mobile networks) to manage mobility, eliminating the need to run simultaneously a set of distinct mobility protocols. Thus, it enables a smoothly transition from IPv4 to IPV6, which has a positive contribution in requirement R8.The IPv4 NAT traversal for Mobile IPv6 is also supported in [94].

## HMIPv6

This section presents and discusses HMIPv6 [18] as a mechanism supported at the network layer to support mobility in a scalable way.

<u>Background</u>

This sub-section discusses in a tutorial perspective the most important mechanisms how HMIPv6 tries to split the mobility support in two parts: mobility inside a specific network domain and mobility among domains. This split is justified by performance and scale reasons, as we following described.

In MIPv6 [12], Binding Updates (Bus) are transferred among the MN, the HA, and the CN, which creates significant additional network overhead, especially in the case the MNs change very frequently their visited networks or the number of MNs roaming increases significantly. Hence, IETF proposed the Hierarchical Mobile IPv6 (HMIPv6) protocol [18] to diminish network overhead due to signaling traffic and optimize the handover performance by separating management of local mobility (i.e. micro-mobility) from global mobility (i.e. macro-mobility), as shown in Figure 7.

HMIPv6 proposes a two-level hierarchical network infrastructure to support host mobility and defines a micro-mobility domain as a group of several visiting networks that a mobile node can potentially visit. Inside the micro-mobility domain, a new entity designated by the Mobility Anchor Point (MAP) is assumed. It is an intermediate unit between local mobility inside the micro-mobility level and global mobility at the macro-mobility level. Since the MAP also acts as a local HA, it captures all packets destined to an MN within its micro-mobility domain and tunnels the captured packets to the MN's current address [31].

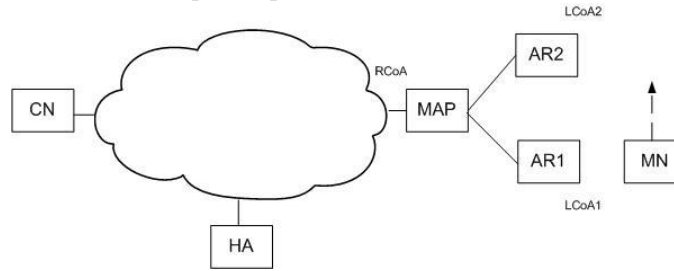

Figure 7 - HMIPv6 Perspective

The MAP also keeps a mapping between the global address (i.e. RCoA) and the local address (i.e. LCoA1). When the MN moves within the same micro-mobility domain but between two distinct visiting networks (e.g. from AR1 to AR2), the MN only updates the new local CoA address (i.e. LCoA2) in the existing mapping table of the MAP. In this way, the local mobility of the node is completely hidden from all the nodes outside the micro-mobility domain associated to the last MAP. Alternatively, as the MN moves to a different micro-mobility domain, the RCoA address also changes because the MAP has also changed, which forces the mobile node to both update the HA and CN with the new RCoA using control messages across the network similarly to those used by the MIPv6 protocol (i.e. macro-mobility).

<u>Critical View</u>

This sub-section discusses (see Table 13) how the hierarchical mobility support in IPv6 can impact some future network requirements already discussed.

Table 13 – HMIPv6 impact on future network requirements

| Functionality | Affected Requirement | Affected Player | Impact Description |
|---|---|---|---|
| Additional tunneling MAP - MN | R3- | All | Increases packet delay |
| High number of MNs inside a domain | R9- | Network Provider | MAP becomes overloaded |
| MAP as a domain controller is like a second HA to MNs | R3- | All | Increases binding update delay that can't meet VoIP delay requirement |

HMIPv6 handles handovers locally through a node designated by MAP (Mobility Anchor Point). Thus, HMIPv6 limits the amount of MIPv6 signaling outside MAP's domain and on average reduces the registration delay. Nevertheless, HMIPv6 could not satisfy the requirements for delay-sensitive traffic such as Voice over IP (VoIP), because HMIPv6 could increase the handover execution delay, originate packet loss and service disruption [13], affecting negatively requirement R3 (Handover efficiency). In addition, for a large number of MNs the MAP could become overloaded, affecting negatively the requirement R9 (Scalability aspect).

As the MN session activity is high and its mobility is relatively low, HMIPv6 may degrade end-to-end data throughput due to the additional packet tunneling between the MAP and the MNs within the domain associated to that MAP. In this way, [31] proposes an Adaptive Route Optimization (ARO) scheme to improve the throughput performance in HMIPv6 networks, using route optimization.

Using HMIPv6, it is required an initial MN authentication against the MAP. For ensuring this, the HMIPv6 specification proposes the use of EAP with Internet Key Exchange (IKE) v2 between the MN and a dedicated AAA server.

## FMIPv6

This section presents and discusses FMIPv6 as a mechanism supported at the network layer to support seamless handovers.

<u>Background</u>

This sub-section discusses in a tutorial perspective the most important mechanisms of FMIPv6 to support mobility in a smooth way.

Fast Handovers for Mobile IPv6 (FMIPv6) [17] is another proposal to optimize MIPv6 [12], shown in Figure 8. FMIPv6 utilizes cooperative Access Routers (ARs) that exchange information with other neighboring ARs that are possible candidates for a MN handover. FMIPv6 also tries to acquire information needed to join a new link before disconnecting the old link. It also configures a bidirectional tunnel between the Previous Access Router (PAR) and the MN at its new COA (nCoA). FMIPv6 requires both PAR and New Access Router (NAR) to buffer traffic during the handover execution. This allows the MN to send packets like it was still connected to its PAR while the MN is finalizing the handover process. Therefore, FMIPv6reduces the handover delay as well as the packet loss. The authors of [10] studied how some link-layer triggers of the 802.21 [9] could potentially enhance the performance of FMIPv6.

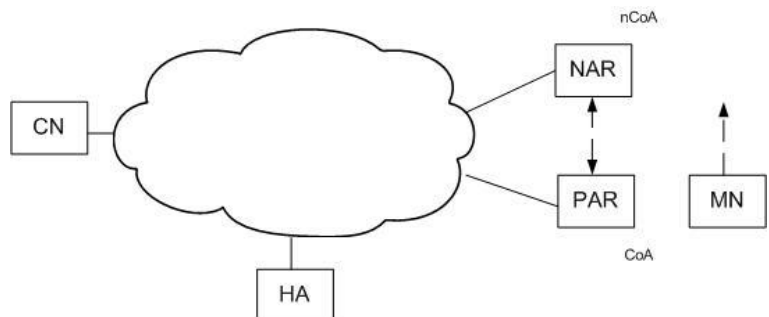

Figure 8 - FMIPv6 Perspective

<u>Critical View</u>

In this sub-section is discussed how FMIPv6 [17] impacts some of the future network requirements already discussed. The aspects discussed in the next paragraphs related with FMIPv6 are summarized in Table 14.

FMIPv6 was proposed to reduce handover latency and to minimize transmission errors during MIPv6 handovers. To fulfil these goals, FMIPv6 proposes new proactive functionalities such as prediction of the direction of terminal movement, anticipated binding update, and anticipated address configuration [15]. These FMIPv6 proactive characteristics, including the routing optimization [92], have a positive contribution to the requirement R3 (Handover efficiency). In addition, FMIPv6 allows an anticipated reaction to movement, by using layer two triggers, followed by the transfer of packets from the current to the next access-router (or NAP). Although FMIPv6 improves MIPv6 performance in terms of handover latency and packet loss, it could not effectively reduce signaling overhead (due to the additional messages exchanged for handover anticipation) and it could cause duplication of data messages circulating in the network. These aspects have a negative impact on R3 (Handover efficiency). Nevertheless, the authors of [13] argue that FMIPv6 does not influence so much the handover latency due to the fact that the major component of this latency is related with the new NAP discovery, which it is optimized by FMIPv6 operation.

Table 14 – FMIPv6 impact on future network requirements

| Functionality | Affected Requirement | Affected Player | Impact Description |
|---|---|---|---|
| MN movement detection | R3+ | All | Decreases handover delay |
| Extra signaling traffic between access routers | R3- | All | Extra overhead and latency |
| Deployment strategy | R8- | Network Provider, terminal | Requires a coordination among network providers |
| Pre-authentication MN-target AP | R5+ | User | Enables MN authentication; Faster resumption of service after roaming |
| Pre-authentication MN-target AP | R3+ | All | Enables MN authentication; Faster resumption of service after roaming |

For a proper FMIPv6 operation, the NAPs, in conjunction with the handover traffic, need to collaborate among them and, transfer contextual information such as access control, QoS, and header compression. In addition, FMIPv6 will not work properly if there are multiple L2 access technologies being used. This aspect has a negative impact on the deployment of this technology (requirement R8). Some research efforts to solve this problem can be found in [39-40]. For all these operations, the NAPs must have the necessary security associations established via mechanisms not yet detailed on the FMIPv6 specification [19]. Finally, the use of a 802.1X pre-authentication between an MN and a target AP while the MN is associated with the previous AP would enable authentication to be carried out (i.e. positive contribution to requirement R5) in advance of the handover finalization, which would allow faster resumption of service after roaming [16] (i.e. also a positive contribution to R3).

## Deployment Analysis and Comparison among Selected Mobility Proposals at Network Layer

This section performs an analysis and comparison on the deployment of all the selected mobile proposals at the network layer. We start with MIPv4. This requires in each network domain the deployment of two mobility agents: Home Agent and Foreign Agent. In opposition, MIPv6 requires only a single agent, Home Agent, in each network domain. Although, the mobility management of MIPv6 is not completely transparent to the CN (if RO is enabled) in opposition to what occurs in the case of MIPv4. In fact, MIPv6 requires a new mobility agent running at CN.

Proxy Mobile IPv6 (PMIPv6) [41] proposes a solution that supports mobility in a complete transparent way to MNs. This design relies on the fact the network is responsible for tracking the movements of the host and eventually initiates the required mobility signaling on the mobile host behalf. In addition, PMIPv6 requires that all mobile access gateways use shared addresses. In this way, a mobile host does not detect any change with respect to its Layer 3 connection, even after it moves from one mobile access gateway to another. So, an extension was proposed to simplify this operational requirement by making a reservation for special addresses that can be used for this purpose [93]. Other interesting work [42] discusses how multihomed hosts can operate in MIPv6.

The MAP node of HMIPv6 is a single point of network failure. In the case a MAP fails, its binding cache content is lost, resulting in loss of communication among mobile and correspondent nodes. This problem may be solved by using multiple MAPs on the same domain and a reliable context transfer protocol among them. However, MAP redundancy is outside the scope of the HMIPv6 specification [18].

The methodology by which Access Routers exchange information about their neighbors could be via Proxy Router Advertisements. But this is out of the scope of FMIPv6 specification [17]. There is an IETF proposal for supporting this, i.e. Candidate Access Router Discovery (CARD) [16], but its use between administrative domains is not recommended, until the policy issues involved are more thoroughly understood. Therefore, it seems that FMIPv6 can be unrealistic from the network operator point of view. In fact, in a real and competitive market, the network operators are unlikely to promote or support handovers to their competitors. One possible solution for this problem is using the Terminal-Controlled Mobility Management proposed in [14], where different access networks may be completely independent of each other, as it does not need any changes in the existing network infrastructure. Another advantage of FMIPv6 is it allows users to have full control of their mobile networking needs [22]. In addition, [10] shows that the behavior of FMIPv6 (whether a reactive or proactive operation is executed) is highly dependent on the timely availability of link layer information. Finally, FMIPv6 operation requires a single HA per network domain.

Some mobility proposals are a mixture of HMIPv6 and FMIPv6 [1] [15] but they have some drawbacks. The proposal of [1] could induce a triangle-type routing problem. Such triangular routing may cause packet delivery delay and a waste of bandwidth [13]. Furthermore, since this combined proposal requires significant network modifications, its deployment could require a long time [1]. The design of Fast Handover for HMIPv6 (F-HMIPv6) was to allow more fast handoff and localized mobility management [15]. However, F-HMIPv6 may show some synchronization issues originating losses because some packets are neither delivered nor buffered. In addition, F-HMIPv6 is a solution with significant complexity with several mobility entities (Home Agents, MAP Agents) and it does not support route optimization.

# CONCLUSION

Assuming a NGN future networking environment, the various design choices of the studied mobility proposals have been classified into three categories, verifying for each proposal if MNs and CNs are (not) mobility-aware [65]. In this way, the first design of this taxonomy (i.e. design 1 – D1) classifies a mobility proposal in which only MNs are mobility-aware. The second design (D2) classifies proposals in which neither CNs nor MNs are mobility-aware. Finally, D3 is applied to mobility solutions where both MNs and CNs are mobility-aware.

The complete set of mobility proposals previously discussed can be now summarized and classified, using the previous taxonomy (please consult Tables 15 and 16 as a guidance to our following discussion).

Table 15 - Qualitative analysis of mobility proposals

| Proposal | 1 | 2 | 3 | 4 | 5 | 6 | 7 | 8 | 9 | 10 | 11 | 12 | 13 | 14 |
|---|---|---|---|---|---|---|---|---|---|---|---|---|---|---|
| MIPv6 | D1 | G | H. Ag. | N | Y | N | N | N | N | N | - | - | - | [12] |
| RO | D3 | G | H. Ag. | Y | Y | N | N | N | N | N | - | - | - | [92] |
| HMIP | D1 | L | H. Ag, MAP | N | Y | N | N | Y | N | N | - | - | - | [18] |
| FMIP | D1 | G | H. Ag. | N | Y | N | N | N | N | Y | + | 0 | + | [17] |
| MCoA | D1 | G | H. Ag. | N | Y | Y | Y | N | Y | N | - | - | - | [27] |
| NEMO | D1 | G | H. Ag. | N | Y | Y(1) | Y(1) | Y | N | N | - | - | - | [19] |
| Multi-Homed | D1 | G | H. Ag. | N | Y | Y | Y(1) | N | Y | N | - | - | - | [42] |
| PMIP | D2 | G | H. Ag. | N | Y | N | N | Y | N | N | - | - | - | [41][93] |
| DSMIP | D1 | G | H. Ag. | Y | Y | N | N | N | N | N | - | - | - | [94] |

Table 16 – Auxiliary information to consult table 15

| ID | Description |
|---|---|
| 1 | Mobility-Aware Entities (D1/D2/D3) |
| 2 | Local (L)/Global (G) Scale? (L/G) |
| 3 | Changes to Architecture? (Modified or new Entities) |
| 4 | Changes to OSI Stack? (Y/N) |
| 5 | Support Legacy Applications? (Y/N) |
| 6 | Support Host Multi-Homing? (Y/N) |
| 7 | Support Flow Mobility? (Y/N) |
| 8 | Energy-Efficient? (Y/N) |
| 9 | Support Load Balancing? (Y/N) |
| 10 | Support Seamless Handover? (Y/N) |
| 11 | Impact on Service/Content Provider (+/0/-) |
| 12 | Impact on Network Provider (+/0/-) |
| 13 | Impact on User/Terminal (+/0/-) |
| 14 | Reference |
| (1) | If MCoA (+flow bindings) extension is also used |
| D1 | CN does not support mobility |
| D2 | CN/MN do not support mobility |
| D3 | Both CN/MN are mobility-aware |

All the studied MIPv6 proposals deployed at Layer 3 (Network) can be classified in terms of their design as D1, meaning MIPv6 protocol and its extensions oblige at least MNs to be involved directly with the mobility management. However, there are two exceptions to what has been concluded in the last phrase. The first exception refers to PMIPv6 extension (D2) which completely alleviates the hosts for managing the mobility. The second exception is about the RO extension (D3) that involves both MNs and CNs.

Focusing the reader's attention to the operating scale of the mobility proposals, all the mobility proposals are global with the exception of HMIP. At this point, there is an interesting tradeoff between optimizing the mobility support and the cost associated to its deployment. In fact, solutions that support local mobility (HMIP) to diminish in some situations the handover latency or to optimize the routing efficiency (MIPv6 - RO) normally imply the deployment of additional mobility entities like agents or controllers in the network infrastructure, increasing not only the operator cost but also the complexity of the network management.

In terms of the deployment strategy, HMIP, NEMO and PMIP have great possibilities to save the energy of the host battery because they attenuate (or remove) from the terminal the burden of managing its mobility.

The analysis among all the various proposals shows that globally MIPv6 with its extensions, on one hand, supports efficiently the diverse mobility requirements, on the other hand, it impacts negatively all the players. A notable exception is FMIPv6, which shows a neutral impact on network providers and a positive impact on both service provider and terminal/user. Nevertheless, FMIPv6 requires coordination among network operators that is normally very difficult to obtain in reality unless there are some good incentives to enable cooperation among them.

Nodes with multiple network interfaces have the great advantage of connecting simultaneously to different networks, and enable their users to enjoy high-performing ubiquitous communication [81]. In this way, the mobility proposals were also analyzed from the point of view of multihoming (multiacess). Apparently only MIPv6 with MCoA and multihoming extensions supports host multiacess. A very interesting open issue to investigate further is how to coordinate multihoming across the OSI layer stack [112], e.g., how LTE carrier aggregation can be coordinated with multihoming extensions of MIPv6.

The host mobility is not a viable solution to efficiently support the distinct traffic types and user preferences in future network environments. In this way, we have checked if our selected mobility proposals can support flow mobility. We have concluded that flow mobility is only supported by MIPv6/NEMO with MCoA extension. In addition, as the number of mobile flows will (exponentially) increase some scalable solutions should be also investigated such as the aggregation of mobile flows, per traffic type or per destination network domain.

Finally, some notorious requisites to be supported by mobile technologies operating in self-organized and intelligent future networking infrastructures are as follows [112]: flow mobility, data offloading, load balancing, and vertical (i.e. from physical to application layer) multihoming.